\documentclass[aps,preprint,prd,nofootinbib]{revtex4}

\usepackage{graphicx}
\usepackage{psfrag}
\usepackage{epsfig}

\begin{document}

\title{An Upgraded Version of the Generator BCVEGPY2.0 for
Hadronic Production of $B_c$ Meson and Its Excited States}
\author{Chao-Hsi Chang$^{1,2}$ \footnote{email:
zhangzx@itp.ac.cn}, Jian-Xiong Wang$^{3}$\footnote{email:
jxwang@mail.ihep.ac.cn} and Xing-Gang Wu$^{2}$\footnote{email:
wuxg@itp.ac.cn}}
\address{$^1$CCAST (World Laboratory), P.O.Box 8730, Beijing 100080,
P.R. China.\\
$^2$Institute of Theoretical Physics, Chinese Academy of Sciences,
P.O.Box 2735, Beijing 100080, P.R. China.\\
$^3$Institute of High Energy Physics, P.O.Box 918(4), Beijing
100049, P.R. China}

\begin{abstract}
An upgraded version of the package [BCVEGPY2.0: Chao-Hsi Chang,
Jian-Xiong Wang and Xing-Gang Wu, Comput. Phys. Commun. {\bf 174}
(2006) 241-251] is presented, which works under LINUX system and is
named as BCVEGPY2.1. Using this version with a GNU C compiler, users
may simulate the $B_c$-events in various experimental environments
very conveniently. It has been rearranged for better modularity and
code reusability (less dependency among various modules) than
BCVEGPY2.0 has. Furthermore, in the upgraded version a special and
convenient executable-file {\bf run} as default is available
according to one's wish, i.e., the file is obtained in the way that
the GNU command {\bf make} compiles the codes requested by precise
purpose with the help of a master {\bf makefile} in the main code
directory. Finally, this paper may also be considered as an erratum
of the original BCVEGPY2.0, i.e., here
the errors (typo mainly) in BCVEGPY2.0 have been corrected.\\


\end{abstract}

\maketitle

\noindent{\bf NEW VERSION PROGRAM (BCVEGPY2.1) SUMMARY}\\

\noindent{\it Title of program} : BCVEGPY2.1\\

\noindent{\it Program obtained from} : CPC Program Library or the
Institute of Theoretical Physics, Chinese Academy of Sciences,
Beijing, P.R. China: {$www.itp.ac.cn/\,\widetilde{}\;zhangzx/bcvegpy2.1$}.\\

\noindent{\it Reference to original program} : BCVEGPY2.0\\

\noindent{\it Reference in CPC} : Comput. Phys. Commun. {\bf 174},
241(2006)\\

\noindent{\it Does the new version supersede the old program}:
No\\

\noindent{\it Computer} : Any LINUX based on PC
with FORTRAN 77 or FORTRAN 90 and GNU C compiler as well.\\

\noindent{\it Operating systems} : LINUX.\\

\noindent{\it Programming language used} : FORTRAN 77/90.\\

\noindent{\it Memory required to execute with typical data} : About
2.0 MB.\\

\noindent{\it No. of bytes in distributed program, (including
PYTHIA6.2)} : About 1.2 MB.\\

\noindent{\it Distribution format} : .tar.gz .\\

\noindent{\it Nature of physical problem} : Hadronic production of
$B_c$ meson itself and its excited states.\\

\noindent{\it Method of solution} : The code with option can
generate weighted and un-weighted events. An interface to PYTHIA
is provided to meet the needs of jets hadronization in the
production.\\

\noindent{\it Restrictions on the complexity of the problem} : The
hadronic production of $(c\bar{b})$-quarkonium in $S$-wave and
$P$-wave states via the mechanism of gluon-gluon fusion are given by
the so-called `complete calculation' approach. \\

\noindent{\it Reasons for new version} : Responding to the
feedback from users, we rearrange the program in a convenient way
and then it can be easily adopted by the users to do the
simulations according to their own experimental environment (e.g.
detector acceptances and experimental cuts). We have paid many
efforts to rearrange the program into several modules with less
dependency among the modules, the main program is slimmed down and
all the further actions are decoupled from the main
program and can be easily called for various purposes. \\

\noindent{\it Typical running time} : The typical running time is
machine and user-parameters dependent. Typically, for production of
the $S$-wave $(c\bar{b})$-quarkonium, when IDWTUP=$1$, it takes
about 20 hour on a 1.8 GHz Intel P4-processor machine to generate
1000 events; however, when IDWTUP=$3$, to generate $10^6$ events it
takes about 40 minutes only. Of the production, the time for the
$P$-wave $(c\bar{b})$-quarkonium will take almost
two times longer than that for its $S$-wave quarkonium.\\


\noindent{\it Summary of the changes (improvements)} :\\

\noindent (1) The structure and organization of the program have
been changed a lot. The new version package BCVEGPY2.1 has been
divided into several modules with less cross communication among the
modules (some old version source files are divided into several
parts for this purpose). The main program is slimmed down and all
the further actions are decoupled from the main program so that they
can be easily called for various applications. All of the Fortran
codes are organized in the main code directory named as {\it
bcvegpy2.1}, which contains the main program, all of its
prerequisite files and subsidiary `folders' (subdirectory to the
main code directory). The method for setting the parameter is the
same as that of the previous versions \cite{bcvegpy1,bcvegpy2}, i.e.
the parameters are set in a file named parameter.F. Each subsidiary
`folders' contains the necessary files to complete specific tasks
accordingly. There are in total seven modules/`folders' in the
program:

\begin{itemize}
\item The module {\bf generate}: it is the key module, which
contains the files for generating the $B_c$ events. There are seven
source files in this `folder': evntinit.F, colorflow.F, genevnt.F,
py6208.F (a nickname of PYTHIA6.208 \cite{pythia}), totfun.F,
outerpdf.F and initmixgrade.F. The function of the module {\bf
generate} is to set the initialize conditions for event simulation;
to establish a connection with PYTHIA \cite{pythia}; to establish a
connection to the parton distribution functions (PDFs) that are not
included in PYTHIA according to specific need; to record the color
flow information of the generated $B_c$ events and may provide it
according to one's need; to calculate the kernel for phase space
integration with the help of {\bf swave} module and {\bf pwave}
module; to do the phase space integration with the help of {\bf
phase} module. A useful trick for generating the mixed type of $B_c$
events is suggested (three types of mixed events are provided in the
generator \cite{bcvegpy2}, e.g. by setting the parameter IMIXTYPE=2,
one can generate the mixed events for the two color-singlet
$(c\bar{b})$-quarkonium states $(c\bar b)_{\bf 1}(^{1}S_{0})$ and
$(c\bar b)_{\bf 1}(^{3}S_{1}) $). The file initmixgrade.F is used to
initialize the importance sampling function for Monte Carlo
simulation, i.e., either by using the importance sampling function
given by the current VEGAS running \cite{vegas} or by using the
existing importance sampling function recorded in an existing grade
file in {\it data} subdirectory that has already been generated by
earlier VEGAS running. Once the importance sampling function has
been obtained by VEGAS, it is recorded in a grade file (with suffix
.grid) automatically, and can be conveniently used (by
initmixgrade.F) for later usage without running VEGAS again. Some
more detail on this point will be shown in the following item (2).

\item The module {\bf phase}: it contains the files for generating
the allowed phase-space point and for generating an importance
sampling function with VEGAS program \cite{vegas}. It contains
three source files: phase$\_$gen.F, phase$\_$point.F and vegas.F.

\item The module {\bf swave}: it contains the files for
calculating the square of the amplitudes for producing the four
color-singlet and color-octet $(c\bar{b})$-quarkonium in $S$-wave:
$(c\bar b)_{\bf 1}(^{1}S_{0})$, $(c\bar b)_{\bf 1}(^{3}S_{1}) $,
$(c\bar b)_{\bf 8}(^{1}S_{0})$ and $(c\bar b)_{\bf 8}(^{3}S_{1})$,
where the subscripts $\bf {1}$ and $\bf{8}$ stand for
color-singlet and color-octet accordingly. Note that in fact the
configurations $(c\bar b)_{\bf 8}(^{1}S_{0})$ and $(c\bar b)_{\bf
8}(^{3}S_{1})$ play comparatively an important role only for the
production of the $P$-wave excited $B_c$ states as shown in
Ref.\cite{changwu}. It contains five source files: s$\_$bound.F,
s$\_$common.F, s$\_$foursets.F, s$\_$free.F and s$\_$samp.F.

\item The module {\bf pwave}: it contains the files for
calculating the square of the amplitudes for producing the four
color-singlet $(c\bar{b})$-quarkonium in $P$-wave:
$(c\bar{b})_{\bf 1}(^1P_1)$ and $(c\bar{b})_{\bf 1}(^3P_J) \,
(J=1,2,3)$. It contains six source files: p$\_$lorentz.F,
p1p1amp.F, pj0amp.F, pj1amp.F, pj2amp.F and p$\_$samp.F.

\item The module {\bf pybook}: it contains the files for
initializing the subroutine PYBOOK of PYTHIA to record the events.
The user may conveniently switch off this module in the main program
to use his/her own way to record the data. It contains five source
files: pybookinit.F, uphistrange.F, uppydump.F, uppyfact.F and
uppyfill.F.

\item The module {\bf setparameter}: it contains the files for inner
use of the parameters (mainly generates some short notations for the
parameters) that have been set in parameter.F. It has only two
source files: simparameter.F and uperror.F, where uperror.F lists
some typical error messages for the cases when the input parameters
are out of allowed (reasonable) range.

\item The module {\bf system}: it contains files to open or
close the record files and to print out certain running messages at
the intermediate steps according to need, which may tell the users
at what step the program is running. It contains six source files:
upopenfile.F, uplogo.F, vegaslogo.F, updatafile.F,
upclosegradefile.F and upclosepyfile.F.
\end{itemize}

\begin{figure}
\centering
\includegraphics[width=0.60\textwidth]{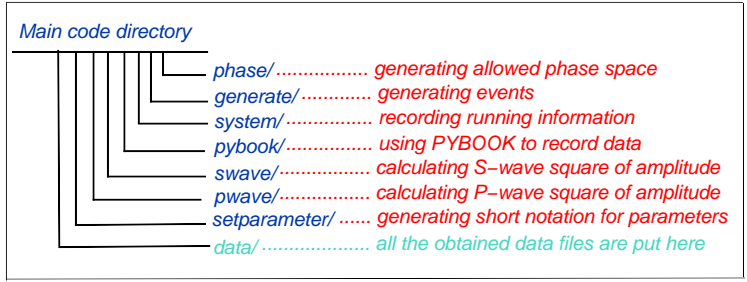}
\caption{The schematic structure for the new version of BCVEGPY.}
\label{struct}
\end{figure}

Each module is equipped with its own {\bf makefile} that will be
used to make a library of the same name, e.g. the {\bf makefile} in
the `folder' {\bf swave/} will be used by the GNU command {\bf make}
to generate a library swave.a. These sub-makefiles are orchestrated
by a master {\bf makefile} in the main code directory. Libraries
required for the main program are listed in the LIBS variable of the
master {\bf makefile} and built automatically by invoking the
sub-makefiles:
\begin{displaymath}
{\rm LIBS = generate.a\;\; swave.a\;\; pwave.a\;\; phase.a\;\;
setparameter.a\;\; system.a\;\; pybook.a }
\end{displaymath}

By running the {\bf make} command under the directory of the main
codes, the master {\bf makefile} is called automatically and the
requested executable-file {\bf run} is obtained. Therefore, the
program acquires good modularity and code reusability, and the users
can easily make the BCVEGPY generator to suit their experimental
simulation environment. Namely, to connect this generator with
his/her own generator package such as ATHENA (used by ATLAS group),
Gauss (used by LHCb group) and SIMUB (used by CMS
group)\footnote{The first version BCVEGPY1.0 has already been
introduced into ATHENA \cite{bcvegpy1,chafik} and SIMUB
\cite{chengm}, and the import of BCVEGPY2.1 into LHCb package Gauss
is in progress \cite{gaoyn}.} only a few pieces of the program need
to be changed. By doing this way, the time for compiling the Fortran
source files can be saved, because once the source file has been
compiled, it does not need to be recompiled again unless some
changes have been made. The schematic structure of the program is
shown in FIG.\ref{struct}. Note that in order to let the {\bf make}
command work smoothly, especially, to deal with some preprocessor
parameters in the source files, all the suffixes of the source files
(with suffix {\bf .for}) in the original version of BCVEGPY2.0
\cite{bcvegpy2} are renamed as {\bf .F} and
the suffixes of the original header files (with suffix {\bf .f}) are renamed as {\bf .h}.\\

\noindent (2) To simulate the events in mixture of the $B_c$ and
its excited states properly and in order to save CPU time, we
offer an additional option in the package BCVEGPY2.1: by setting
IVEGASOPEN=0 and IGRADE=1, one may generate the mixed $B_c$ events
just by reading the existent importance sampling function and
total cross-section from `previous' VEGAS runs for the production,
when they have been saved in date files with initmixgrade.F
accordingly. Thus it means that when using the package with this
option, one runs VEGAS only once. Whereas, in BCVEGPY2.0
\cite{bcvegpy2}, there is only one option by setting IVEGASOPEN=1:
it is to run VEGAS every time, although the importance sampling
function obtained by each run is recorded in an existed grade file
in the {\it data} subdirectory. For the new option offered here,
the trick to save CPU time is not to run VEGAS, once the
importance sampling function has
been recorded in the {\bf .grid} file already.\\

\noindent (3) In the package BCVEGPY2.1, for testing the importance
sampling function obtained by VEGAS and simulating the true
situation as well as possible, we have further offered another
option: with setting IVEGASOPEN=0 and IGRADE=0, VEGAS or the
importance sampling function (grade function) generated in previous
runs will not be used at all. In this case, being different from the
other options, all of the files with suffix {\bf .grid} must be
generated and saved via previous run(s) for the states which one
would like to mix according to their weights. Namely one should have
them in advance, because the program needs to read the total
cross-sections of these states from the {\bf .grid} files
accordingly, so as to determine the relative weight of each state
in making the mixing correctly. \\

\noindent (4) As stated in Ref.\cite{bcvegpy2}, the precision of the
generated importance sampling function by running VEGAS can be
improved by properly adjusting the maximum iteration number, the
number of calls to the integrand in each iteration, and the number
of bins (number of sub-intervals dividing the [0,1] interval) as
well. In the present version, we define three overall parameters:
NVEGITMX, NVEGCALL and NVEGBIN in the head file invegas.h, with
these three parameters one can easily adjust them. Similarly, for
convenience, the parameters which are not changed frequently are
also defined in the source file run.F, e.g. NUMOFEVENTS (number of
events to be generated), ENERGYOFTEVA
(TEVATRON energy) and ENERGYOFLHC (LHC energy).\\

\noindent (5) All the files containing the results are put in the
subdirectory {\bf data}. For clarity, in recording the obtained
data, all of the grade files are ended with the suffix {\bf
.grid}, all the intermediate files, that record the used parameter
values and the VEGAS running information, are ended with suffix
{\bf .cs} and all the files that record the differential
distributions, e.g. the transverse momentum and rapidity
distributions of the $(c\bar{b})$-quarkonium, are ended with
suffix {\bf .dat}. In a specific grade file, not only the
information for the sampling importance function is recorded, but
also the total cross-section and the maximum differential
cross-section
are recorded.\\

\noindent (6) A simple script, named as {\bf do} which does all
the necessary jobs for generating the events, is put in the main
code directory. For convenience, a script, named as {\bf pnuglot}
\footnote{This script is programmed based on the {\bf gnuplot}
with a lot of plotting parameters pre-set to reasonable
values.\cite{formcalc}} that may produce a high-quality plot in
Encapsulatetd PostScript (EPS) format from a data file, is
supplied here (taken from the FormCalc
package \cite{formcalc}).\\

\noindent (7) We have found several typos in BCVEGPY2.0
\cite{bcvegpy2}, so we list all of them and their corrections here:

\begin{itemize}
\item In the subprogram bcvegpy.for: in line 493-507, the values
for MSTU(112) and PARU(112) should be exchanged; in line 212,
IMIXTYPE=1 should be replaced by IMIXTYPE=3; in line 956,
PETA=PYP(I,19) should be replaced with PSETA=PYP(I,19).

\item In the subprogram genevnt.for: in line 505, ICOLUP(1,3)=503
should be replaced by ICOLUP(1,4)=503.

\item In the subprogram parameter.for: we should add the statement
`double complex colmat, bundamp' for assigning the nature of the
variables in all of the three subroutines there; in the subroutine
setparameter when calling the subroutine setctq6, to match to the
default setting of the parameters, the original setctq6(3) should
be replaced by setctq6(4), which corresponds to the default
setting of PDF as CTEQ6L1 \cite{6lcteq}.
\end{itemize}

\begin{center}

{\bf ACKNOWLEDGEMENTS}
\end{center}
The authors would like to thank Prof. Y.N. Gao, Dr. Z.W. Yang and
Dr. J.B. He for helpful suggestions on improving the program. This
work was supported by the Natural Science Foundation of China. The
author (X.G. Wu) would like to thank the support from the China
Postdoctoral Science Foundation. \\


\begin{thebibliography}{s2}

\bibitem{bcvegpy1} Chao-Hsi Chang, Chafik Driouich, Paula Eerola and Xing-Gang Wu,
Comput. Phys. Commun. {\bf 159}, 192 (2004); hep-ph/0309120.

\bibitem{bcvegpy2} Chao-Hsi Chang, Jian-Xiong Wang and Xing-Gang
Wu, Comput. Phys. Commun. {\bf 174}, 241 (2006); hep-ph/0504017.

\bibitem{pythia} T. Sjostrand, Comput. Phys. Commun. {\bf 82}, 74 (1994).

\bibitem{vegas} G.P. Lepage, J. Comp. Phys {\bf 27}, 192 (1978).

\bibitem{changwu} Chao-Hsi Chang, Cong-Feng Qiao, Jian-Xiong Wang and
Xing-Gang Wu, Phys. Rev. D{\bf 71}, 074012(2005).

\bibitem{6lcteq} H.L. Lai, et al., hep-ph/0201195; JHEP {\bf 0207}, 012
(2002).

\bibitem{formcalc} T. Hahn and M. Rauch, hep-ph/0601248(2006).

\bibitem{chafik} private communication with Paula
Eerola and Chafik Driouichi.

\bibitem{chengm} private communication with A.A. Belkov, G.M. Chen,
S. Shulga and S.H. Zhang; S.H. Zhang, A.A. Belkov, S. Shulga and
Guo-Ming Chen, Chin. Phys. Lett. {\bf 21}, 2380(2004).

\bibitem{gaoyn} private communication with Y.N. Gao, J.B. He, P. Robbe
and Z.W. Yang.

\end{thebibliography}
\end{document}